\documentclass[prd,preprint,showpacs,eqsecnum,nofootinbib]{revtex4}
\usepackage{bm}
\usepackage{stmaryrd}
\usepackage{amsmath}
\usepackage{amssymb}
\usepackage{graphicx}
\usepackage{textcomp}
\usepackage{calrsfs}
\usepackage{yfonts}

\newcommand{\nn}{\nonumber}

\newcommand{\be}{\begin{equation}}
\newcommand{\ee}{\end{equation}}
\newcommand{\ben}{\begin{equation*}}
\newcommand{\een}{\end{equation*}}
\newcommand{\bea}{\begin{eqnarray}}
\newcommand{\eea}{\end{eqnarray}}

\newcommand{\bnabla}{\bm{\nabla}}
\newcommand{\bGamma}{\bm{\Gamma}}

\DeclareMathOperator{\Tr}{Tr}
\DeclareMathOperator{\tr}{tr}

\begin{document}

\title{Casimir-Polder repulsion:
Polarizable atoms, cylinders, spheres, and ellipsoids
}

\date{\today}

\author{Kimball A. Milton}\email{milton@nhn.ou.edu}
\author{Prachi Parashar}\email{prachi@nhn.ou.edu}
\author{Nima Pourtolami}\email{nimap@ou.edu}
\affiliation{Homer L. Dodge Department of
Physics and Astronomy, University of Oklahoma, Norman, OK 73019-2061, USA}
\author{Iver Brevik}\email{iver.h.brevik@ntnu.no}
\affiliation{Department of Energy and Process Engineering, 
Norwegian University
of Science and Technology, N-7491 Trondheim, Norway}

\begin{abstract}
Recently, the topic of Casimir repulsion has received a great deal of
attention, largely because of the possibility of technological application.
The general subject has a long history, going back to the self-repulsion
of a conducting spherical shell and the repulsion between a perfect electric
conductor and a perfect magnetic conductor.  Recently it has been observed
that repulsion can be achieved between ordinary conducting bodies,
provided sufficient anisotropy is present.  For example, an anisotropic
polarizable atom can be repelled near an aperture in a conducting plate.
Here we provide new examples of this effect, including the repulsion on such
an atom moving on a trajectory nonintersecting a conducting cylinder; in
contrast, such repulsion does not occur outside a sphere.  Classically,
repulsion does occur between a conducting ellipsoid placed in a uniform
electric field and an electric dipole. The Casimir-Polder force between
an anisotropic atom and an anisotropic dielectric semispace does not
 exhibit repulsion.  The general systematics of repulsion
are becoming clear.
\end{abstract}

\maketitle
\section{Introduction}
Although known since the time of Lifshitz's work on the subject
\cite{lifshitz61}, repulsive Casimir forces have recently received
serious scrutiny \cite{Levin:2010zz}.  Experimental confirmation
of the repulsion that occurs when dielectric surfaces are separated
by a liquid with an intermediate value of the dielectric constant
has appeared \cite{capasso09}, although this seems devoid of much
practical application. The context of our work is the
considerable interest in utilizing the quantum vacuum force or
the Casimir effect in nanotechnology employing mesoscopic objects
\cite{Rodriguez:2010zz}.

The first repulsive Casimir stress in vacuum was found by Boyer
\cite{boyer68}, who discovered the still surprising fact that
the Casimir self-energy of a perfectly conducting spherical
shell is positive.  (This has become somewhat less mysterious,
since the phenomenon is part of a general pattern
\cite{bender,milton,abalo1,abalo2}.)  Boyer later observed that
a perfect electrical conductor and a perfect magnetic conductor
repel \cite{boyer74}, but this also seems beyond reach, since the unusual
electrical properties must be exhibited over a wide frequency range.  The
analogous effect for metamaterials also seem impracticable \cite{McCauley}.

Thus it was a significant advance when Levin et al.\
showed examples of repulsion between conducting objects, in particular
between an elongated cylinder above a conducting plane with a circular
aperture \cite{Levin:2010zz}.  
(See also Ref.~\cite{maghrebi}.)
They computed the quantum vacuum forces between conducting objects, by using
  impressive numerical finite-difference time-domain and
boundary-element methods.

We subsequently showed \cite{Milton:2011ni} that repulsive Casimir-Polder
forces between anisotropic atoms and a conducting half-plane, and even
between such an atom and a conducting wedge of rather large opening angle,
could be achieved.  Of course, we must be careful to explain what we mean
by repulsion: the total force on the atom is attractive, but the component
of the force perpendicular to the symmetry axis of the conductor changes
sign when the atom is sufficiently close to that axis.  This is the
only component that survives in the case of an aperture in a plane, so
our analytic calculation provided a counterpart to the numerical work
of Ref.~\cite{Levin:2010zz}.  

In this paper we give some further examples.  After demonstrating, in
Sec.~\ref{sec2}, that Casimir-Polder repulsion
between two atoms requires that both be sufficiently
anisotropic, we show in Sec.~\ref{sec3}
 that the force between one such atom and a conducting
cylinder is repulsive for motion confined to a perpendicular line not intersecting with
the cylinder, provided the line is sufficiently far from the cylinder.
The analogous effect does not occur for a spherical conductor
(Sec.~\ref{sec4}), as one
might suspect since at large distances such a sphere looks like an isotropic
atom.  The classical interaction between a dipole and a conducting ellipsoid
polarized by an external field is examined in Sec.~\ref{sec5}, which, as
expected, yields a repulsive region. In contrast,  in Sec.~\ref{sec6}, we examine the 
Casimir-Polder interaction
of an anisotropic atom with an anisotropic dielectric half-space, but this fails
to reveal any repulsive regime.

In this paper we set $\hbar=c=1$, and all results are expressed in Gaussian
units except that Heaviside-Lorentz units are used for Green's dyadics.

\section{Casimir-Polder repulsion between atoms}
\label{sec2}
\begin{figure}
\begin{center}
\includegraphics{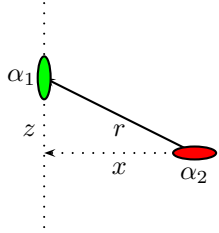}
\caption{\label{pol-atoms} Casimir-Polder interaction between two atoms
of polarizability $\bm{\alpha}_1$ and $\bm{\alpha}_2$ separated by a distance $r$.
Atom 1 is predominantly polarizable in the $z$ direction, while atom 2
is predominantly polarizable in the $x$ direction.  The force on atom 1
in the $z$ direction becomes repulsive sufficiently close to the polarization
axis of atom 2 provided both atoms are sufficiently anisotropic.}
\end{center}
\end{figure}

The interaction between two polarizable atoms, described by general
polarizabilities $\bm{\alpha}_{1,2}$, with the relative separation
vector given by  $\mathbf{r}$ is \cite{CP1,CP2}
\be
U_{\rm CP}=-\frac1{4\pi r^7}\left[\frac{13}2\Tr\bm{\alpha}_1\cdot
\bm{\alpha}_2-28\Tr(\bm{\alpha}_1\cdot\mathbf{\hat r})(\bm{\alpha}_2
\cdot\mathbf{\hat r})+\frac{63}2(\mathbf{\hat r}\cdot\bm{\alpha}_1
\cdot\mathbf{\hat r})(\mathbf{\hat r}\cdot\bm{\alpha}_2
\cdot\mathbf{\hat r})\right].\label{generalcp}
\ee
This formula is easily rederived by the multiple scattering
technique as explained in Ref.~\cite{brevikfest}.
This reduces, in the isotropic case, $\bm{\alpha}_i=\alpha_i\bm{1}$,
to the usual Casimir-Polder (CP) energy, 
$U_{\rm CP}=-\frac{23}{4\pi r^7}\alpha_1\alpha_2$.
Suppose the two atoms are only polarizable in perpendicular directions,
$\bm{\alpha}_1=\alpha_1\mathbf{\hat z \hat z}$,
$\bm{\alpha}_2=\alpha_2\mathbf{\hat x \hat x}$.  Choose atom 2 to be at the origin.
The configuration is shown in Fig.~\ref{pol-atoms}.
Then, in terms of the polar angle $\cos\theta=z/r$, the $z$-component of
the force on atom 1 is
\be
F_z=-\frac{63}{8\pi}\frac{\alpha_1\alpha_2}{x^8}\sin^{10}\theta\cos\theta
(9-11\sin^2\theta).
\ee
In this paper, we are considering motion for fixed $x=r\sin\theta$,
in the $y=0$ plane.
Evidently, the force is attractive at large distances, vanishing
as $\theta\to0$, but it must change sign at small values of $z$ for fixed $x$,
since the energy also vanishes as $\theta\to\pi/2$.  The force component
in the $z$ direction vanishes when $\sin\theta=3/\sqrt{11}$ or $\theta=1.130$
or 25$^\circ$ from the $x$ axis.\footnote{After the first version of this paper
was prepared, Ref.~\cite{SS} appeared, which rederived these results, and then
went on to extend the calculation to Casimir-Polder repulsion by an anisotropic
dilute dielectric sheet with a circular aperture.  The authors quite correctly
point out that the statement in Ref.~\cite{Milton:2011ni}  that no repulsion is possible in
the weak-coupling regime is erroneous.}

No repulsion occurs if one of the atoms is isotropically polarizable.
If both have cylindrically symmetric anisotropies, but with respect
to perpendicular axes,
\be
\bm{\alpha}_1=(1-\gamma_1)\alpha_1\mathbf{\hat z \hat z}+\gamma_1\alpha_1\bm{1},\quad 
\bm{\alpha}_2=(1-\gamma_2)\alpha_2\mathbf{\hat x \hat x}+\gamma_2\alpha_2\bm{1},
\ee
it is easy to check that if both are sufficiently anisotropic repulsion will occur.
For example, if $\gamma_1=\gamma_2$ repulsion in the $z$ direction will take place close to the plane
$z=0$ if $\gamma\le0.26$.

\section{Repulsion of an atom by a conducting cylinder}
\label{sec3}

Now we turn to the Casimir-Polder (CP) interaction between a polarizable
body (``atom'') and a macroscopic body.  That interaction is generally given by
\be
E_{\rm CP}=-\int_{-\infty}^\infty d\zeta\tr \bm{\alpha}\cdot\bm{\Gamma}(\mathbf{r,r}),
\ee
where $\mathbf{r}$ is the position of the atom and $\zeta$ is
the imaginary frequency,
in terms of the polarizability of the atom $\bm{\alpha}$ and
the Green's dyadic due to the macroscopic body, which for
a body characterized by a permittivity $\varepsilon$
satisfies the differential equation
\be
\left(\frac1{\omega^2}\bnabla\times\bnabla\times
-\bm{1}\varepsilon(\mathbf{r})\right)\cdot\bGamma(\mathbf{
r,r'})=\bm{1}\delta(\mathbf{r-r'}).\label{gdif}
\ee  In this paper, except for Sec.~\ref{sec6}, 
 we will consider perfect conducting boundaries $S$ 
immersed in vacuum, in which case
we need to solve this equation with $\varepsilon=1$ for $\bGamma$
subject to the boundary conditions
$\mathbf{\hat n}\times \bGamma(\mathbf{r,r'})\bigg|_{\mathbf{r}\in S}=0$,
where $\mathbf{\hat n}$ is the normal to the surface of the conductor,
which just states that the tangential components of the
 electric field must vanish on the conductor.

\begin{figure}
\begin{center}
\includegraphics{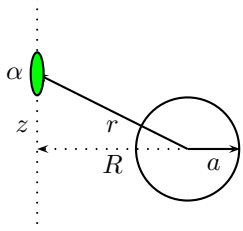}
\caption{\label{fig-cyl-atom} Interaction between an anisotropically polarizable
atom and a conducting cylinder of radius $a$.  The force on the atom along a line
which does not intersect the cylinder is considered.  If the atom is only
polarizable in that direction, and the line lies sufficiently far from the
cylinder, the force component along the line changes sign near the point of
closest approach.}
\end{center}
\end{figure}

Let us henceforth assume that the polarizability has negligible frequency dependence
(static approximation), and, in order to maximize the repulsive effect, the atom
is only polarizable in the $z$ direction, the direction of the trajectory
(assumed not to intersect the cylinder),
 in which case the quantity we need to compute for 
a conducting cylinder of radius $a$ is given by \cite{bezerra}
\bea
\int_{-\infty}^\infty \frac{d\zeta}{2\pi}\Gamma_{zz}(r,\theta)&=
&\sum_{m=-\infty}^\infty
\int_0^\infty\frac{d\kappa}{(2\pi)^3}\frac\pi{2a}
\frac1{K_m(\kappa a)K'_m(\kappa a)}
\bigg\{\frac{m^2}{r^2}K_m^2(\kappa r)+\kappa^2K_m^{\prime2}(\kappa r)\nn\\
&&\quad\mbox{}-\cos2\theta \kappa a[I_m(\kappa a)K_m(\kappa a)]'
\left(-\frac{m^2}{r^2}
K_m^2(\kappa r)+\kappa^2K_m^{\prime 2}(\kappa r)\right)\bigg\}.
\eea
The geometry we are considering is illustrated in Fig.~\ref{fig-cyl-atom}.
It gives greater insight to give the transverse electric (TE) 
and transverse magnetic (TM) contributions to the CP energy:
\begin{subequations}
\bea
E^{\rm TE}_{\rm CP}&=&-\frac{\alpha_{zz}}{4\pi}\sum_{m=-\infty}^\infty
\int_0^\infty d\kappa\,\kappa
\frac{I'_m(\kappa a)}{K'_m(\kappa a)}\left[\frac{\cos^2\theta}{r^2} m^2K_m^2
(\kappa r)
+\kappa^2\sin^2\theta K_m^{\prime 2}(\kappa r)\right],\\
E^{\rm TM}_{\rm CP}&=&\frac{\alpha_{zz}}{4\pi}\sum_{m=-\infty}^\infty
\int_0^\infty d\kappa\,\kappa
\frac{I_m(\kappa a)}{K_m(\kappa a)}\left[\frac{\sin^2\theta}{r^2} m^2K_m^2
(\kappa r)+\kappa^2\cos^2\theta K_m^{\prime 2}(\kappa r)\right].
\eea
\end{subequations}
  The distance of the atom from the center of the
 cylinder is $r=R/\sin\theta$, where $R$ is the distance of
closest approach and $\theta$ is the polar angle, which ranges from
0 when the atom is at infinity to $\pi/2$ when the atom is closest to
the cylinder.

\begin{figure}
\begin{center}
\includegraphics{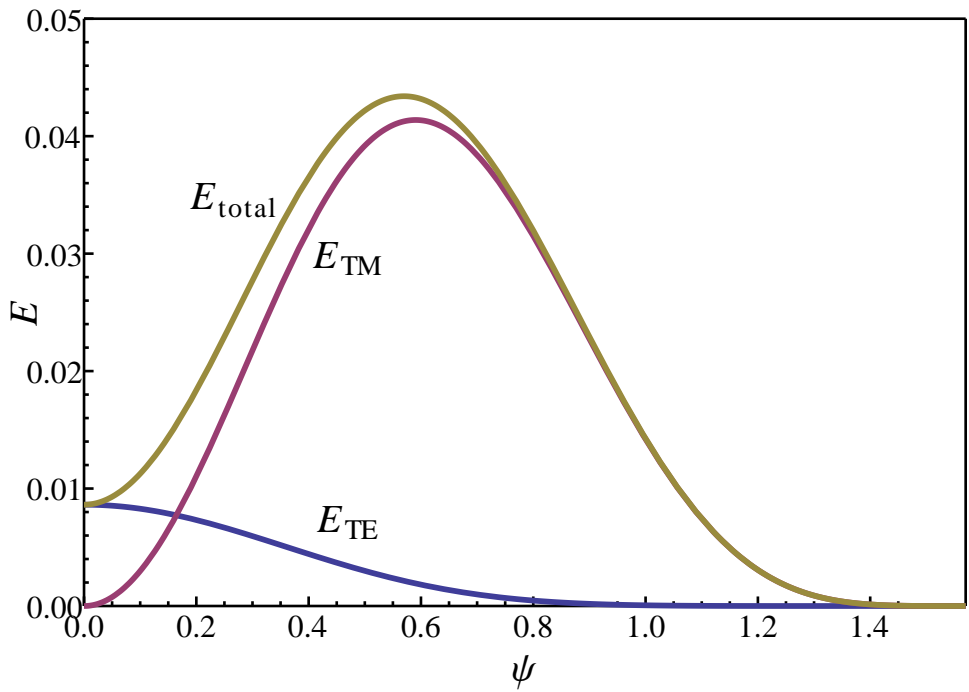}
\caption{\label{fig1} $m=0$ contributions to the Casimir-Polder energy
between an anisotropic atom and a conducting cylinder.  The (generally) lowest curve
(blue) is the TE contribution, the second (magenta) is the TM contribution, and the top
curve (yellow) is the total CP energy.  In this case, the distance of closest
approach of the atom is taken to be 10 times the radius of the cylinder.
The energy $E$ is plotted as a function of $\psi=\pi/2-\theta$.}
\end{center}
\end{figure}

\begin{figure}
\begin{center}
\includegraphics{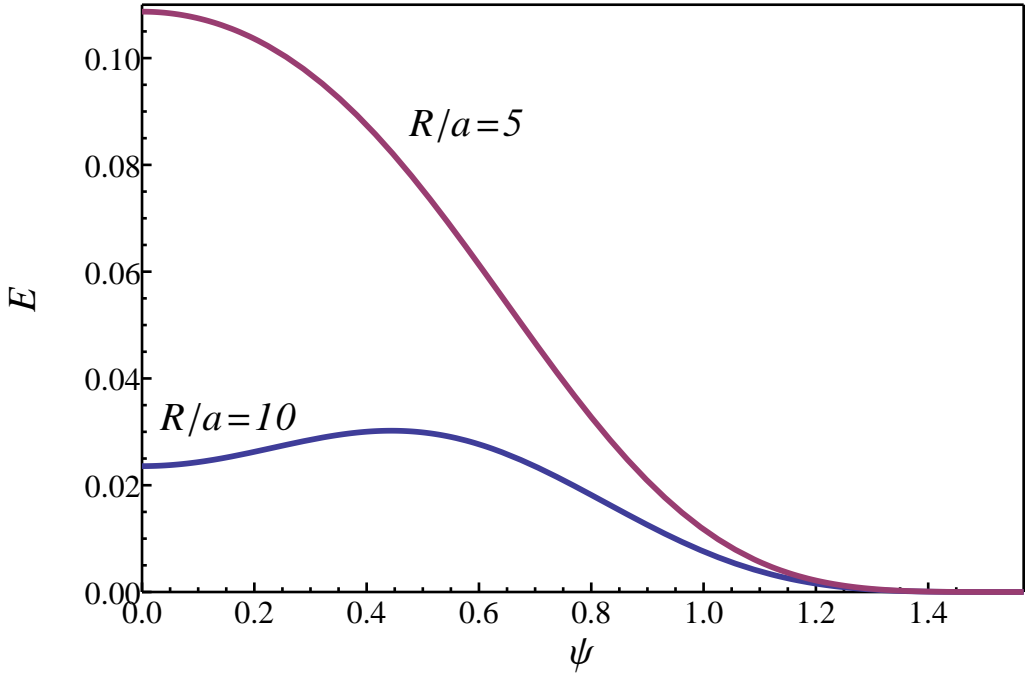}
\caption{\label{fig2} The CP energy between an anisotropic atom
and a conducting cylinder.  Plotted is the total CP energy, the upper curve 
for the distance of closest approach $R$ being 5 times the cylinder
radius $a$, the lower curve for the distance of closest approach 10 times the
radius.  The curves move up slightly as more $m$ terms are included,
but have completely converged by the time $m=3$ is included.
Repulsion is clearly observed when $R/a=10$, but not for $R/a=5$.}
\end{center}
\end{figure}

At large distances, the CP force is dominated by the $m=0$ term in the
energy sum.  Figure \ref{fig1} shows that for $m=0$ the TM mode dominates
except near the position of closest approach, where only the TE mode
is nonzero.  This indicates that there is a region of repulsion near
$\theta=\pi/2$, since the total energy has a minimum for small
$\psi=\pi/2-\theta$. This effect is partially washed out by including higher $m$ modes,
as seen in Fig.~\ref{fig2}, which shows the effect of including the first 5
$m$ values.  But the repulsion goes away if the line of motion passes
too close to the cylinder.  Numerically, we have found that to have
repulsion close to the plane of closest approach requires that
$a/R<0.15$.

\section{CP interaction between atom and conducting sphere}
\label{sec4}

It is straightforward to derive the TE and TM contributions 
for the interaction between a completely anisotropic
 atom and a conducting sphere as
\begin{subequations}
\bea
E^{\rm TM}&=&\frac{\alpha_{zz}}{2\pi R^4}\cos^4\theta\sum_{l=1}^\infty
(2l+1)\int_0^\infty dx \,g_l(x),\\
E^{\rm TE}&=&\frac{\alpha_{zz}}{4\pi R^4}\cos^6\theta\sum_{l=1}^\infty
(2l+1)\int_0^\infty dx \,f_l(x),
\eea
\end{subequations}
where 
\begin{subequations}
\bea
g_l(x)&=&x\frac{s_l'(xa\cos\theta/R)}{e_l'(xa\cos\theta/R)}\left[\frac12\cos^2
\theta e_l^{\prime2}(x)+\frac{l(l+1)\sin^2\theta e_l^2(x)}{x^2}\right],\\
f_l(x)&=&x\frac{s_l(xa\cos\theta/R)}{e_l(xa\cos\theta/R)}e_l^2(x),
\eea
\end{subequations}
where the modified Riccati-Bessel functions are
\be
s_l(x)=\sqrt{\frac{\pi x}2}I_{l+1/2}(x),\quad
e_l(x)=\sqrt{\frac{2 x}\pi}K_{l+1/2}(x).\quad
\ee

We expect in the case of a sphere not to see Casimir repulsion at large
distances.  The reason is that far from the sphere it appears to be an
isotropic atom, which, as we have seen above will not give a
repulsive force on another completely anisotropic atom.  Indeed, far
from the sphere we can replace the Bessel functions of argument $xa/r$
by their leading small argument approximations and we easily find
\begin{subequations}
\be
E^{\rm TM}\sim \frac{\alpha_{zz} a^3}{4\pi r^7}(13+7\sin^2\theta),
\quad a/r\to 0.\label{TMcp}
\ee
The TE mode contributes
\be
E^{\rm TE}\sim \frac{\alpha_{zz}a^3}{4\pi r^7}\frac74\cos^2\theta,
\quad a/r\to0.\label{cpte}
\ee
\end{subequations}
We see here the expected isotropic electric
polarizability of a conducting sphere
$\alpha_{{\rm sp},E}=\bm{1} a^3$.
We note that the TM result (\ref{TMcp}) coincides with the result obtained
from Eq.~(\ref{generalcp}).  The TE contribution is, in fact, the coupling
between the electric polarizability of the atom and the magnetic polarizability
of the sphere
$\alpha_{{\rm sp},M}=-\frac{a^3}2\bm{1}$ \cite{embook}.

To see this, we first remind the reader of the CP interaction between
isotropic atoms possessing both electric and magnetic polarizabilities
\cite{feinberg},
\be
U_{\rm CP}=-\frac{23}{4\pi r^7}(\alpha_1^E\alpha_2^E+\alpha_1^M\alpha_2^M)
+\frac7{4\pi r^7}(\alpha_1^E\alpha_2^M+\alpha_1^M\alpha_2^E).\label{totalcp}
\ee
When the atoms are not isotropic it is easy to deduce the generalization
of this, using the methods described in
Ref.~\cite{brevikfest}, starting from the 
multiple-scattering coupling term between electric 
and magnetic dyadics,
\be
E_{\rm em}=-\frac{i}2\Tr\ln\left(1+\bm{\Phi}_0 \mathbf{T}_1^E\cdot\bm{\Phi}_0
\mathbf{T}_2^M\right)\approx
-\frac{i}2\Tr\bm{\Phi}_0\cdot\mathbf{V}_1^E\bm{\Phi}_0^e\cdot \mathbf{V}_2^M,
\ee
where the last form reflects weak coupling, and we are considering the
interaction between one object having purely electric susceptibility and
a second object having purely magnetic susceptibility, so
\be
\mathbf{V}_1^E=4\pi \bm{\alpha}^E_1\delta(\mathbf{r-r_1}),\quad
\mathbf{V}_2^M=4\pi \bm{\alpha}^M_2\delta(\mathbf{r-r_2}).
\ee
This formula is expressed in terms of the magnetic Green's dyadic,
\be
\bm{\Phi}_0=-\frac{\zeta^2}{4\pi R^3}\mathbf{R}\times(|\zeta| R+1)e^{-|\zeta|R}.
\ee
Then, an immediate calculation yields the electric-magnetic CP interaction
\be
U_{\rm CP,EM}=\frac7{8\pi R^7}\tr (\mathbf{\hat R}\times\bm{\alpha}^E)
(\mathbf{\hat R}\times\bm{\alpha}^M),
\ee
which indeed for isotropic polarizabilities
gives the second term in Eq.~(\ref{totalcp}).  The result
(\ref{cpte}) is now an immediate consequence for a conducting sphere
interacting with an atom only polarizable in the $z$ direction.

Evidently, no repulsion can occur in this CP limit where the conducting sphere
is regarded as an 
isotropically polarizable atom.  In fact, numerical evaluation
shows no repulsion occurs at any separation distance between the
sphere and the atom.

\section{ Electrostatic force between a  conducting ellipsoid and a dipole}
\label{sec5}
In this section we return, for heuristic reasons, to the electrostatic situation of the
interaction between a fixed dipole and a conducting body.  Such have been given
considerable attention lately \cite{Levin:2010zz,Milton:2011ni,lj}.  Here we 
consider the interaction between a perfectly conducting ellipsoid polarized by a
constant electric field and a fixed dipole.  The polarization of the ellipsoid
by the dipole is neglected at this stage.  This is a much simpler calculation than
the more interesting one of the interaction between a dipole and a ellipsoid, but we
justify the inclusion of the details of the simpler calculation here because it allows
us to approach the 
complexity of the full calculation.  Elsewhere, we will present that calculation and
the corresponding quantum Casimir-Polder calculation, building on the work of
Ref.~\cite{Graham:2011ta}.
\subsection{Ellipsoidal coordinates}

Consider first a conducting uncharged solid ellipsoid with semiaxes $a>b>c$, centered at the origin $x=y=z=0$. The   semiaxis $c$ lies along the $z$ axis. The electrostatic potential $\phi$ in the external region can be described in terms of ellipsoidal coordinates
 $\xi, \eta, \zeta$, corresponding to solutions for $u$ of the cubic equation
\begin{equation}
\frac{x^2}{a^2+u}+\frac{y^2}{b^2+u}+\frac{z^2}{c^2+u}=1. \label{1}
\end{equation}
The coordinate intervals are in general
\begin{equation}
\infty >\xi \geq -c^2, \quad -c^2 \geq \eta \geq -b^2, \quad -b^2\geq \zeta \geq -a^2. \label{2}
\end{equation}
We will henceforth assume axial symmetry around the $z$ axis. In that case, $b \rightarrow a, \,\zeta \rightarrow -a^2 $, and the ellipsoidal coordinates $\xi, \eta, \zeta$ reduce to oblate spheroidal coordinates $\xi$ and $\eta$ restricted to the intervals
\begin{equation}
\infty > \xi \geq -c^2, \quad -c^2 \geq \eta \geq -a^2. \label{3}
\end{equation}
If $\rho=\sqrt{x^2+y^2}$ denotes the horizontal radius in the  plane $z=$ constant, the cubic equation (\ref{1}) reduces to the quadratic equation
\begin{equation}
u^2-(\rho^2-a^2-c^2+z^2)u-(\rho^2-a^2)c^2-z^2a^2=0 \label{4}
\end{equation}
for $u=(\xi, \eta)$. The solution for $u=\xi$ corresponds to the positive square root:
\begin{equation}
\xi=\frac{1}{2}(\rho^2-a^2-c^2+z^2)+\frac{1}{2}\sqrt{(\rho^2-a^2+c^2)^2+z^2(2\rho^2+2a^2-2c^2+z^2)}. \label{5}
\end{equation}
At the surface of the ellipsoid, $\xi=0$, whereas in the external region, $\xi>0$. Note that in the $xy$ plane ($z=0$) the expression for $\xi$  simplifies to  $\xi=\rho^2-a^2$, when
$\rho>a$. The solution for $u=\eta$ corresponds to the same expression (\ref{5}) but with the negative square root.

Surfaces of constant $\xi$ and $\eta$ are oblate spheroids and hyperboloids of revolution, the surfaces intersecting orthogonally. On the symmetry axis $\rho=0$ one has $\xi=-c^2+z^2, \, \eta=-a^2$. The relations between $\xi, \eta$ and $z, \rho$ are
\begin{equation}
z=\pm \sqrt{ \frac{(\xi+c^2)(\eta+c^2)}{c^2-a^2}}, \quad \rho=\sqrt{\frac{(\xi+a^2)(\eta+a^2)}{a^2-c^2}}. \label{6}
\end{equation}
We will henceforth be concerned with the half-space $z\geq 0$ only.

\subsection{Ellipsoid situated in a uniform electric field}

Assume now that the ellipsoid is placed in a uniform electric field ${\bf E}_0$, directed along the $z$ axis. We take the electrostatic potential $\phi$ to be zero on the ellipsoid surface. With quantities $R_\xi$ and $R_\eta$ defined as
\begin{equation}
R_\xi=(\xi+a^2)\sqrt{\xi+c^2}, \quad R_\eta=(\eta+a^2)\sqrt{\eta+c^2}, \label{7}
\end{equation}
 the Laplace equation in the external region $\xi \geq 0$ can be written as
 \begin{equation}
\nabla^2 \phi \equiv \frac{4}{\xi-\eta}\left[ \frac{R_\xi}{\xi+a^2}\frac{\partial}{\partial\xi}\left(R_\xi\frac{\partial 
\phi}{\partial \xi}\right)- \frac{R_\eta}{\eta+a^2} \frac{\partial}{\partial \eta} \left(R_\eta \frac{\partial 
\phi}{\partial \eta}\right)\right]=0. \label{8}
 \end{equation}
 The potential due solely to ${\bf E}_0$ is
 \begin{equation}
 \phi_0=-E_0 z, \label{9}
 \end{equation}
 and we write the full potential $\phi$ in the form
 \begin{equation}
 \phi=\phi_0[1+F(\xi)], \label{10}
 \end{equation}
 so that $\phi_0F$ denotes the modification due to the ellipsoid. The boundary condition at the surface is $F(0)=-1$.

 Inserting Eq.~(\ref{10}) into Eq.~(\ref{8}) we find the following equation for $F$,
\begin{equation}
\frac{d^2 F}{d\xi^2}+\frac{dF}{d\xi}\frac{d}{d\xi}\ln \left[ R_\xi (\xi+c^2)\right]=0. \label{11}
\end{equation}
The solution can be written as
\begin{equation}
\phi=\phi_0\left[ 1-\frac{\int_\xi^\infty \frac{ds}{(s+c^2)R_s}}{\int_0^\infty \frac{ds}{(s+c^2)R_s}}  \right].
 \label{12}
\end{equation}

We can also express the solution in terms of the incomplete beta function, 
defined as
\begin{equation}
B_x(\alpha, \beta)=\int_0^x t^{\alpha-1}(1-t)^{\beta-1} dt. \label{13}
\end{equation}
Some manipulation yields
\begin{equation}
\int_\xi^\infty \frac{ds}{(s+c^2)R_s}=\frac{1}{(a^2-c^2)^{3/2}}
B_{(a^2-c^2)/(\xi+a^2)}\left(\frac{3}{2}, -\frac{1}{2}\right), \label{14}
\end{equation}
and so we can write the  final answer for the potential as
\begin{equation}
\phi=\phi_0\left[ 1-\frac{B_{(a^2-c^2)/(\xi+a^2)}\left(
\frac{3}{2}, -\frac{1}{2}\right)}{B_{1-c^2/a^2}\left(
\frac{3}{2}, -\frac{1}{2}\right)} \right]. \label{15}
\end{equation}
For small values of $x$ the following expansion may be useful,
\begin{equation}
B_x(\alpha, \beta)=\frac{x^\alpha}{\alpha}(1-x)^\beta \left[1+
\sum_{n=0}^\infty \frac{B(\alpha+1,n+1)}{B(\alpha+\beta,n+1)}x^{n+1}\right], 
\label{16}
\end{equation}
where $B(\alpha,\beta)=\Gamma(\alpha)\Gamma(\beta)/\Gamma(\alpha+\beta)$ 
is the complete beta function. In our case, the limit $x \ll 1$ 
corresponds to the minor semiaxis $c$ being only slightly less than the 
major semiaxis $a$.

In the following, we shall need the expression for the $z$ component of the 
electric field, $E_z=-\partial \phi/\partial z$, at an arbitrary point 
$(\rho,z)$ in the exterior region. It is here convenient first to 
differentiate the relation (\ref{4})  ($u=\xi$) with respect to $z$, 
keeping $\rho$ constant, to obtain
\begin{equation}
\left(\frac{\partial \xi}{\partial z}\right)_\rho=\frac{2(\xi+a^2)}
{\xi-\eta}\sqrt{\frac{(\xi+c^2)
(\eta+c^2)}{c^2-a^2}}. \label{17}
\end{equation}
With $x=(a^2-c^2)/(\xi+a^2)$ we have
\begin{equation}
\frac{\partial B_x \left(\frac{3}{2},-\frac{1}{2}\right)}
{\partial z}=\frac{\partial \xi}{\partial z}\,\frac{\partial x}
{\partial \xi}\,\frac{\partial B_x\left(\frac{3}{2},-\frac{1}{2}\right)}
{\partial x}=2\frac{(a^2-c^2)}{(\xi+c^2)(\xi-\eta)}(-\eta-c^2)^{1/2}. 
\label{18}
\end{equation}
Then, from Eq.~(\ref{15}),
\begin{equation}
E_z=E_0\left[ 1-\frac{B_{(a^2-c^2)/(\xi+a^2)}\left(\frac{3}{2},-\frac{1}{2}
\right)}{B_{1-c^2/a^2}\left(\frac{3}{2}, -\frac{1}{2}\right)}
-\frac{2(a^2-c^2)^{1/2}(\xi+c^2)^{-1/2}(\eta+c^2)}{B_{1-c^2/a^2}
\left(\frac{3}{2}, -\frac{1}{2}\right)}\frac1{\xi-\eta} \right]. \label{19}
\end{equation}
For large values of $z$ and arbitrary $\rho$ the influence from the 
ellipsoid must evidently fade away, $E_z \rightarrow E_0$.

In the $xy$ plane where $z=0, \xi+a^2=\rho^2$, $\eta+c^2=0$, we have
\begin{equation}
E_z(z=0)=E_0\left[ 1-\frac{B_{(a^2-c^2)/\rho^2}\left(\frac{3}{2},
-\frac{1}{2}\right)}{B_{1-c^2/a^2}\left(\frac{3}{2}, -\frac{1}{2}\right)} 
\right]. \label{20}
\end{equation}
When $\rho=a$ (on the surface), $E_z(z=0)=0$ as it should.

\subsection{Force on a dipole}

Assume now that a dipole ${\bf p}=p_z \mathbf{\hat z}$ 
is situated at rest in the position $(\rho, z)$. 
The dipole is taken to be polarized in the $z$ 
direction only. The value of $z$ $( \geq 0)$ is arbitrary, 
whereas the value of $\rho$ is assumed constant. Thus, 
writing $\rho=a+L$, $L$ is the constant horizontal distance between 
the dipole and the edge of the ellipsoid. The force $F_z$ on the dipole is
\begin{equation}
F_z=\nabla_z({\bf p\cdot E}) =p_z\frac{\partial E_z}{\partial z}. \label{21}
\end{equation}
Note that we are ignoring the polarization of the ellipsoid by the field
of the dipole; the ellipsoid acquires a dipole moment only because of the
applied external field.
We thus have to differentiate the expression (\ref{19}) with respect to $z$. 
Performing the calculation along the same lines as above, we obtain
\bea F_z&=&\frac{6p_zE_0}{B_{1-c^2/a^2}\left( \frac{3}{2},-\frac{1}{2}\right)}
\frac{(a^2-c^2)\sqrt{-\eta-c^2}}{(\xi+c^2)(\xi-\eta)} \nn\\
&&\quad\times \left[ 1-\frac{(\xi+a^2)(-\eta-c^2)}{(a^2-c^2)(\xi-\eta)}   
+\frac23\frac{(\xi+c^2)(\eta+c^2)(\xi+\eta+2a^2)}{(a^2-c^2)(\xi-\eta)^2}
   \right]. \label{22}
\eea
At $z=0$, the force vanishes as it should, since $\eta+c^2=0$ then.

Note that the force vanishes if $c/a\to0$, that is, for a disk, because
the integral representing the
incomplete beta function diverges in the limit. 
(It is not to be interpreted as its analytic continuation.) This is not surprising,
for in the limit of a disk, the electric field is just $\mathbf{E}_0$, the
applied constant field.  This is because inserting a perfectly conducting
sheet perpendicular to the field line has no effect on the boundary
conditions.  See also the discussion in Chap.~4 of Ref.~\cite{radbook}.

As a small check, we consider the limit of a sphere, $c^2\to a^2$.
Then, according to Eq.~(\ref{16}), we have
\be
B_{1-c^2/a^2}\left(\frac32,-\frac12\right)\to\frac23a^{-3}(a^2-c^2)^{3/2},
\ee
and
\be
\xi\approx \rho^2+z^2-c^2, \quad \eta=-c^2-\frac{\delta^2 z^2}{\rho^2+z^2},
\ee
in terms of the ultimately vanishing quantity $\delta^2=a^2-c^2$.
Then we immediately obtain
\be
F_z=3 p_z E_0\frac{a^3z}{(\rho^2+z^2)^{7/2}}(3\rho^2-2z^2).\label{dipsh}
\ee
This result also follows immediately from the dipole-dipole interaction
energy
\be
U=-\frac1{r^5}(3\mathbf{r\cdot p}_1\,\mathbf{r\cdot p}_2-r^2 \mathbf{p}_1\cdot
\mathbf{p}_2),
\ee
when we take
\be \mathbf{p}_1=p_z\mathbf{\hat z},\quad \mathbf{p}_2=a^3 E_0\mathbf{\hat z}.
\ee
The force on the sphere (\ref{dipsh}) is attractive at large distance,
because the dipoles become essentially coaxial then, and repulsive at
small distance, because the case of parallel dipoles in a plane is approached
in that situation.

The same features hold for a general ellipsoid.  For short distances,
$z^2\ll \rho^2-a^2+c^2$, we have
\be
\xi=\rho^2-a^2+O(z^2),\quad \eta=-c^2-\frac{z^2(a^2-c^2)}{\rho^2-a^2+c^2}+
O(z^4),
\ee
and then the force is repulsive,
\be
z\to 0:\quad 
F_z=\frac{6 p_z E_0}{B_{1-c^2/a^2}(3/2,-1/2)}\frac{z(a^2-c^2)^{3/2}}{(
\rho^2-a^2+c^2)^{5/2}},
\ee
which reduces in the spherical case to
\be
c\to a:\quad F_z=\frac{9p_zE_0 a^3 z}{\rho^5},
\ee
which agrees with Eq.~(\ref{dipsh}).  And in the large distance limit,
where $\xi\approx z^2$, $\eta\approx-a^2$, the force in general is attractive,
\be
z\to\infty:\quad F_z=-\frac{4 p_z E_0(a^2-c^2)^{3/2}}{B_{1-c^2/a^2}(3/2,-1/2)}
\frac1{z^4},
\ee
which again has the expected limit,
\be
c\to a:\quad F_z=-\frac{6p_zE_0 a^3}{z^4}.
\ee

\section{Interaction of anisotropic atom with anisotropic dielectric}
\label{sec6}
In view of the considerations of Sec.~\ref{sec2}, we might hope that
repulsion could be achieved if an anisotropic atom were placed above
an anisotropic dielectric medium.  Consider such an atom, with polarizability
only in the $z$ direction, $\bm{\alpha}=\alpha\mathbf{\hat z\hat z}$, 
a distance
$a$ above a dielectric with different permittivities in the $z$ direction and
the transverse directions,
\be
\bm{\varepsilon}=\mbox{diag}(\varepsilon_\perp,\varepsilon_\perp,\varepsilon_\|).
\ee
We will assume (see below) that $\varepsilon_\perp$, $\varepsilon_\|>1$.
The Casimir-Polder interaction is
\be
E_{\rm CP}=-\alpha\int_{-\infty}^\infty d\zeta\left(\Gamma_{zz}-\Gamma^0_{zz}\right)
(\mathbf{R,R}),
\ee
where the atom is located at $\mathbf{R}=(0,0,a)$.
Here we have subtracted the free-space contribution.
We can write the Green's dyadic in terms of a transverse Fourier transform,
\be
\bm{\Gamma}(\mathbf{r,r'})=\int\frac{(d\mathbf{k}_\perp)}{(2\pi)^2}e^{i\mathbf{
k_\perp\cdot(r-r')_\perp}}\bm{\gamma}(z,z'),
\ee
where (assuming that $\mathbf{k}_\perp$ lies in the $+x$
direction)
\be
\bm{\gamma}(z,z')=\left(\begin{array}{ccc}
\frac1{\varepsilon_\perp}\frac\partial{\partial z}
\frac1{\varepsilon_\perp'}\frac\partial{\partial z'} g^H&0&
\frac{ik_\perp}{\varepsilon_\perp\varepsilon_\|'}\frac\partial{\partial z}g^H\\
0&-\zeta^2g^E&0\\
-\frac{ik_\perp}{\varepsilon'_\perp\varepsilon_\|}\frac\partial{\partial z'}g^H&
0&\frac{k_\perp^2}{\varepsilon_\|\varepsilon_\|'} g^H\end{array}\right).
\ee
We have followed Ref.~\cite{Schwinger:1977pa}
and used the notation $\varepsilon=\varepsilon(z)$, $\varepsilon'=
\varepsilon(z')$.  Here we have omitted $\delta$-function terms that
do not contribute in the point-splitting limit.  The transverse electric
and transverse magnetic Green's functions satisfy the differential
equations
\begin{subequations}
\bea
\left(-\frac{\partial^2}{\partial z^2}+k_\perp^2-\omega^2\varepsilon_\perp
\right)
g^E(z,z')=\delta(z-z'),\\
\left(-\frac{\partial}{\partial z}\frac1{\varepsilon_\perp}
\frac\partial{\partial z}
+\frac{k_\perp^2}{\varepsilon_\|}-\omega^2\right)
g^H(z,z')=\delta(z-z').
\eea
\end{subequations}

It is rather straightforward to solve these equations and find the Casimir-Polder
energy:
\be
E_{\rm CP}=\frac{\alpha}{4\pi^2}\int_{-\infty}^\infty d\zeta\int (d\mathbf{k_\perp})
\frac{k_\perp^2}{2\kappa}\frac{\bar\kappa-\kappa}{\bar\kappa+\kappa}e^{-2\kappa a},
\label{aniso-atom-diel}
\ee
where $\kappa^2=k_\perp^2-\omega^2$, $\bar\kappa=
\sqrt{(k_\perp^2-\omega^2\varepsilon_\|)/\varepsilon_\perp\varepsilon_\|}$.
Checks of this result are the following: 
\be
\varepsilon_\perp\to\infty: \quad E_{\rm CP}\to -\frac\alpha{8\pi a^4},
\ee
one-third of the usual Casimir-Polder interaction of an isotropic atom with a perfect
conducting plate.  This is what we would have for such an anisotropic atom above
a isotropic conducting plate, because taking $\varepsilon_\perp\to \infty$ imposes
the usual boundary condition that the tangential components of $\mathbf{E}$ vanish
on the surface.
In the other limit, we have no such simple correspondence,
\be
\varepsilon_\|\to\infty: \quad E_{\rm CP}\to \frac\alpha{8\pi a^4}
\left(1+\frac32\sqrt{\varepsilon_\perp}-3\varepsilon_\perp+3\sqrt{\varepsilon_\perp}(\varepsilon_\perp-1)
\ln\frac{\sqrt{\varepsilon_\perp}+1}{\sqrt{\varepsilon_\perp}}\right),
\ee
where the quantity in parentheses varies between $-1/2$ for $\varepsilon_\perp=1$
and $-1$ as $\varepsilon_\perp\to \infty$.

We can check that in all cases, if we ignore
dispersion, Eq.~(\ref{aniso-atom-diel})
yields an attractive result:  $E_{\rm CP}$ scales like $a^{-4}$ times
a numerical integral which is always negative because $\bar\kappa^2-\kappa^2<0$.
Repulsion does not occur in this case because there is no breaking
of translational invariance in the transverse direction.

In fact, 
the electromagnetic force density in an anisotropic nonmagnetic medium is
(see Ref.~\cite{brevikrev}, Eq.~(1.2a))
\be
{\bf f}=-\frac1{8\pi}E_iE_k\nabla \varepsilon_{ik}.
\ee
Assume that the single air-medium interface is flat, lying in the $xy$ plane. 
Then the only nonvanishing component of the gradient $\nabla \varepsilon_{ik}$ 
is the vertical component $\partial_z \varepsilon_{ik}$. If the principal coordinate axes
for $\varepsilon_{ij}$ coincide with the $x$, $y$, $z$ axes, 
then the surface force density $\int f_z \, dz$ (which is subsequently to be
integrated across the surface $z=0$), is directed upwards, because 
$\varepsilon_{\parallel, \perp} >1$. The surface force acts in the direction of the optically thinner medium. 
Now, momentum conservation of the total system asserts that the force on a dipole above the surface acts 
in the downward direction. The dipole force has to be attractive.

That $\varepsilon >1$ for an isotropic medium is a  thermodynamical result. 
For an anisotropic medium, oriented such that the coordinate axes fall together 
with the crystallographic axes, one must analogously have $\varepsilon_{\parallel, \perp}>1$. 
See,  for instance, Sec.~14 in Ref.~\cite{landau-lifshitz}.

Note the contrast with the force on a dipole outside a dielectric wedge, studied in Ref.~\cite{Milton:2011ni}. 
In the latter case, the normal surface force on the inclined (lower) surface necessarily has a vertical 
($z$) component that is downward directed. Momentum conservation for the total system thus no longer 
forbids the force on the dipole to be repulsive.

\section{Conclusions}

Earlier, we observed that Casimir-Polder repulsion along a direction 
perpendicular to the
symmetry axis of a semi-infinite planar conductor or a conducting wedge
and an anisotropically polarizable atom could be achieved in the region
close to the conductor \cite{Milton:2011ni}.  Here we have shown that anisotropically
polarizable atoms can also repel in this sense, provided they are sufficiently
anisotropic, and have perpendicular principal axes.  We further show that 
such an
atom may be repelled by a conducting cylinder, provided,
at closest approach, it is sufficiently far
away from the cylinder, whereas no such phenomenon occurs for a sphere and an
anisotropic atom.  We further discuss a new example of classical
repulsion by considering a polarized ellipsoid interacting with a dipole.
On the other hand, a system of an anisotropically polarizable atom interacting 
via fluctuation forces with an anisotropic dielectric half-space
does not exhibit repulsion.  Apparently, spatial anisotropy is also required
for repulsion between electric bodies. 

\acknowledgments
We thank the US Department of Energy, and the US National Science Foundation,
for partial support of this research.  The support of the ESF Casimir Network
is also acknowledged. 
P.P. acknowledges the hospitality of the
University of Zaragoza.  We thank E. K. Abalo for collaborative discussions.
We also thank K. V. Shajesh and M. Schaden for useful correspondence.

\end{document}